\documentclass[superscriptaddress,showpacs,aps,twocolumn,prb,floatfix]{revtex4}
\pdfoutput=1
\usepackage{amssymb}
\usepackage{amsmath}
\usepackage[dvips]{graphicx}
\usepackage{bm}

\usepackage{color,ulem}

\begin{document}
\title{Non-Fermi liquid behaviour in non-equilibrium transport through Co doped Au chains
connected to four-fold symmetric leads}

\author{S. Di Napoli}
\affiliation{Dpto de F\'{\i}sica, Centro At\'{o}mico Constituyentes, Comisi\'{o}n
Nacional de Energ\'{\i}a At\'{o}mica, Buenos Aires, Argentina}

\author{P. Roura-Bas}
\affiliation{Dpto de F\'{\i}sica, Centro At\'{o}mico Constituyentes, Comisi\'{o}n
Nacional de Energ\'{\i}a At\'{o}mica, Buenos Aires, Argentina}

\author{Andreas Weichselbaum}
\affiliation{Ludwig-Maximilians-Universit\"{a}t München, 80333 Munich, Germany}

\author{A. A. Aligia}
\affiliation{Centro At\'{o}mico Bariloche and Instituto Balseiro, Comisi\'{o}n Nacional
de Energ\'{\i}a At\'{o}mica, 8400 Bariloche, Argentina}

\begin{abstract}
We calculate the differential conductance as a function of temperature and bias voltage, $G(T,V)$, 
through Au monoatomic chains with a substitutional Co atom as
a magnetic impurity, connected to a four-fold symmetric lead. 
The system was recently proposed as a possible scenario for observation of 
the overscreened Kondo physics. Stretching the chain, the system could be tuned through
a quantum critical point (QCP) with three different regimes, overscreened, underscreened and 
non Kondo phases. 
We present calculations of the impurity spectral function by using the 
numerical renormalization group (NRG) for the three different regimes characterizing the QCP. Non trivial
behaviour of the spectral function is reported near the QCP. 
Comparison with results using the non crossing approximation (NCA), shows that the latter
is reliable in the overscreened regime, when the anisotropy is larger than the Kondo temperature.
For these parameters, which correspond to realistic previous estimates,  $G(T,V)$ calculated within NCA
exhibits clear signatures of the non-Fermi liquid behaviour within the overscreened regime. 

\end{abstract}


\maketitle

\section{Introduction}

The Fermi liquid (FL), or the Landau-Fermi liquid theory, is on the basis of our 
understanding of many properties of the metal state at sufficient low temperatures. 
For instance, the electrons in a normal, non-superconducting,  
metal at low temperatures behave as a FL. Also, magnetic impurities with spin $S_I=1/2$ 
embedded in a non-magnetic metal exhibit the Kondo anomaly, that could be theoretically explained 
by the one-channel Kondo (1CK) \cite{kondo} and one-channel
Anderson (1CA) \cite{anderson} models. Both models 
lead to a ground state that can be described by a FL. 
Even for low dimensional systems, such as quantum dots 
coupled to metallic leads, transport measurements at low temperatures were found in 
agreement with FL predictions. Specifically, conductance through a 
single-electron transistor at low temperature is in
quantitative agreement with 
the calculated one from the Anderson impurity model \cite{gg-1ch}.

On the contrary, the non-Fermi liquid (NFL) 
paradigm
describes a system which displays a breakdown of 
the Fermi-liquid properties. A large class of heavy fermions materials, such as Ce and U 
alloys are examples of a metallic state that is not a Fermi-liquid \cite{cox-sawadowski, 
cox-kim}. Exotic properties of these alloys at low temperature, such as significant 
residual entropy and non-saturated magnetic susceptibility can be understood on the basis 
of the two-channel Kondo (2CK) model introduced early by Nozi\`{e}res and Blandin 
\cite{nozieres-blandin}, which is one example of the NFL quantum impurity model. 
In low dimensions, the simplest example of NFL is the Luttinger liquid, 
given by interacting fermions in one dimension \cite{schulz}.  

From the experimental point of view, the realization of the two-channel (2C) state
was studied in a double dot system proposed by Oreg and 
Goldhaber-Gordon,\cite{oreg,potok}, who showed that
the differential conductance as a function of bias voltage $V$ follows a $\sqrt{V}$ behaviour, 
which is, again, characteristic of 2C physics,\cite{potok}.
Much theoretical work have been done, \cite{pustilnik,serge}, in order to develop a theory 
of such experimental setup on the basis of the two channel Anderson 
(2CA) Hamiltonian.  
In this model, two symmetric independent electron modes screen a localized level with 
spin $S_I=1/2$. Among other interesting properties coming up from this model, both, 
the impurity contribution to the entropy at zero temperature, $S=\frac{1}{2} \ln(2)$, 
and the conductance per channel at low-temperatures, $G(T)\simeq a - b\sqrt{T}$, 
 display a NFL behaviour,\cite{bethe,zar,mit}.
The key property to observe the NFL signatures in the above mentioned experiment was the 
setup capability to control the coupling constants between the dot and two independent
reservoirs, $J_1, J_2$, to make them symmetric ones, $J_1 \sim J_2$. 
The requirement of symmetry between the scattering channels,\cite{nozieres-blandin}, is 
very difficult to achieve in real materials, making the NFL observation hard to find. 

Recently, two different realizations of a 2CK effect, with a robust symmetry between the 
two conduction channels have been proposed \cite{tsvelik, solange-1}. In Ref. \onlinecite{tsvelik}, 
Tsvelik \textit{et al.} show  a possible realization of 
the overscreened multi-channel Kondo model in a system of spin chains. In this model, 
$N$ spin $S=1/2$ Heisenberg chains interact with a cluster of $N$ extra spins $1/2$. Some 
interesting examples of real materials that could exhibit the necessary symmetry between the 
scattering channels 
were proposed \cite{tsvelik}.
In Ref. \onlinecite{solange-1} a four-fold symmetric 
Co doped Au chain was proposed as a scenario
to exhibit NFL behaviour by Di Napoli \textit{et al.}. 
The Co atom, considered as a magnetic impurity, mix the $3d^7$ 
and $3d^8$ configurations through the hopping with $5d_{xz}$ and $5d_{yz}$ electrons of Au, 
that play the role of two independent and symmetric scattering channels. The broken axial 
symmetry along the chain by a four-fold symmetric crystal field is an essential ingredient 
to observe the NFL signatures. Stretching the system might be a way to pass through a 
quantum critical point (QCP) that divide three different phases, overscreened, underscreened 
and finally another one without Kondo effect. Specifically, within the overscreened regime, 
two different properties of the NFL behaviour have been found: the conductance per channel 
as a function of temperature through the Co following a $G(T)\simeq a - b\sqrt{T}$ form 
and the Co entropy contribution having a residual value of $S=\frac{1}{2} \ln(2)$ at zero temperature.  

The purpose of the present contribution is to complement the study of the Co-Au chain model with
properties not shown in Ref. \onlinecite{solange-1}. First, we present calculations of the impurity 
spectral function obtained with the numerical renormalization group (NRG)\cite{NRGrefs}
for the three different 
phases reinforcing the presence of a QCP. Non trivial behaviour of the 
spectral density is discussed near to the QCP. Second, we present calculations of the differential 
conductance, $G(T,V)$, through the Co atom within a non-equilibrium situation by using the non 
crossing approximation (NCA) solution of the model in its overscreened regime. We obtain that $G(T,V)$ also 
displays clear signatures of NFL behaviour in this regime. 
We also show the limitations of the NCA to describe the other regimes and to capture very small energy scales.

The paper is organized as follows. 
In section \ref{model} we introduce the model. The 
different regimes of the system are also discussed. In section 
\ref{numerical-results} the numerical 
solution of the model is presented within the NRG and NCA approaches. Finally, in section 
\ref{conclusions} some
conclusions are drown.

\section{Model}
\label{model}

\begin{figure}[tbp]
\includegraphics[width=7cm]{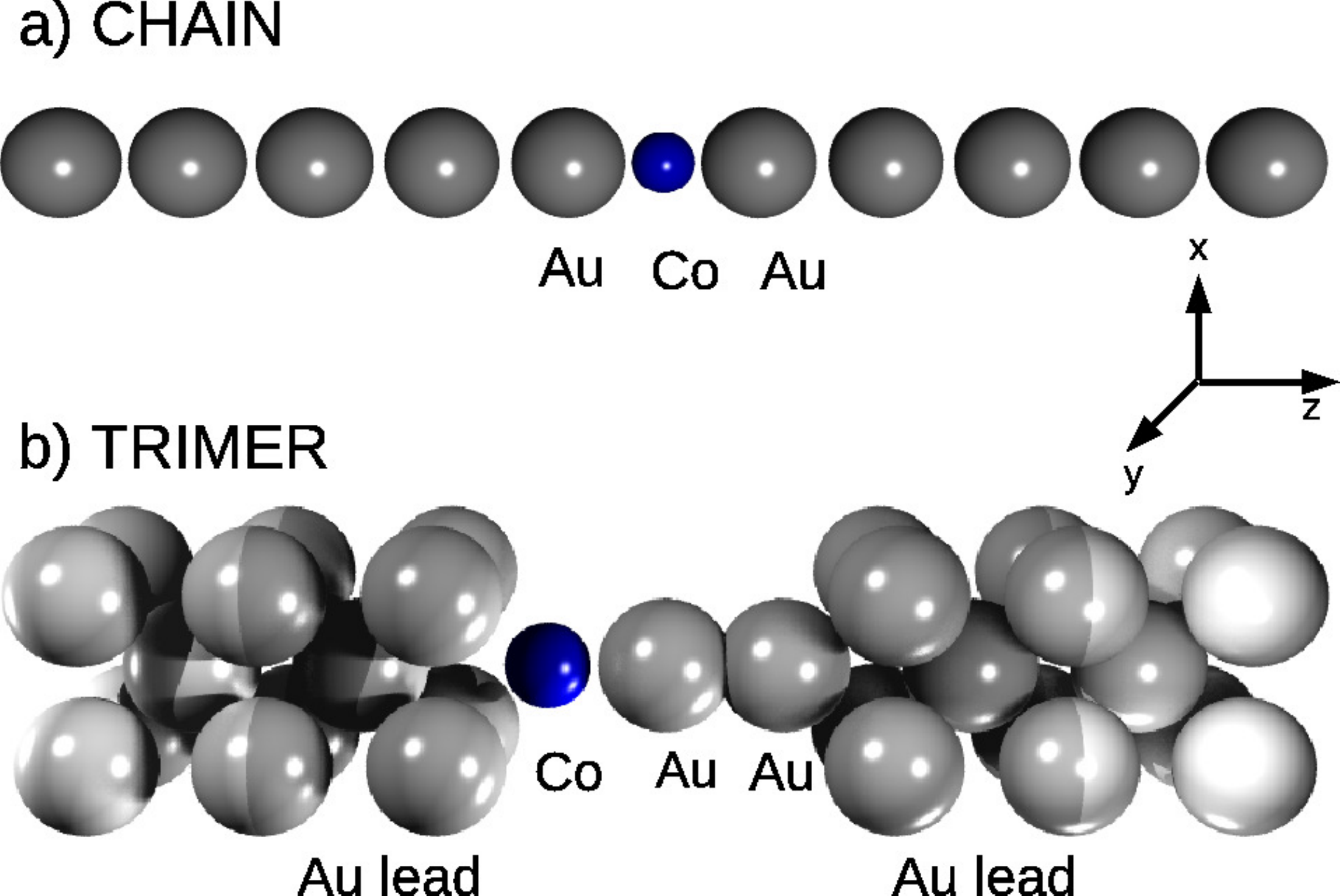}
\caption{ Sketch of the pure chain with a substitutional Co (a) and a trimer connected to a four-fold 
symmetric axis leads (b).}
\label{setup}
\end{figure}

According to \textit{ab-initio} calculations, the Co atom embedded in a Au chain, Fig. \ref{setup}a,
is in a $3d^7$
configuration with the three holes coupled to a total spin $S=3/2$ \cite{solange-1}. This atomic
configuration seems to be robust even if the noble metal changes from Au to Ag and Cu \cite{solange-2}. 
One \textit{d-}hole is shared between the half filled $3d_{xy}$ and $3d_{x^2-y^2}$ ($\Delta_4$-symmetry) 
orbitals while the 
other two are in the degenerate $3d_{xz}$, $3d_{yz}$ ($\Delta_3$-symmetry) ones. 

According to transport measurements, 
the pure Au chain only has $6s$ bands crossing the Fermi level \cite{au-pure}. However, the presence
of environmental O impurities can push up and stabilize the $5d_{xz}$ and $5d_{yz}$ bands of 
Au at and above the Fermi 
level \cite{solange-2}.  While the $\Delta_3$ symmetries of Au chain can be
tuned in order to make these band conductors, there is no similar mechanism to get electrons of Au 
with $\Delta_4$ symmetry at the Fermi level due to their high localization. 
Details of the first principle calculations can be found in 
Refs.\cite{solange-1, solange-2}.

According to this, the $5d_{xz}$ and $5d_{yz}$ bands of Au represent two independent and symmetric channels
that screen the localized moment at the corresponding symmetries of the Co impurity. The $\Delta_4$-symmetry 
levels in Co have frozen its charge and spin fluctuations due to the absence of hybridization with the Au neighbours. 
 
As it was previously mentioned in the introduction, in a system with a four-fold symmetric axis such as a trimer 
with one Co atom connected to body-centered-cubic leads, Fig. \ref{setup}b, the degeneracy between the 
$3d_{xy}$ and 
$3d_{x^2-y^2}$
orbitals of Co is broken localizing the hole in the $3d_{xy}$ orbital. 
Note that exactly the same phyisics is obtained if both $\Delta_4$ orbitals are interchanged.
The energies of the localized three holes in the Co atom 
are $E_{xy}=-0.2$ eV and $E_{xz}=E_{yz}=-0.3$ eV,
where we set the Fermi level as the zero of energy.
The spin-orbit coupling (SOC) in the Co atom induces
a splitting $D$ between the projections $S_z = \pm 3/2$ and $S_z = \pm 1/2$ of the quadruplet that 
belongs to the total
spin $S=3/2$. The calculation of $D$ was exactly done by solving a 120 x 120 matrix of the Hamiltonian of the 
$3d^7$ configurations \cite{kroll}. For the real parameters of the setup shown in Fig. \ref{setup}b 
the value of the anisotropy was found to be $D=1.7$ meV.
While the \textit{ab-initio} calculations \cite{solange-1} have been done in the system represented 
 in Fig. \ref{setup}b, we want to emphasize that the same physics is expected for any length of the 
 Au chain between the leads.

With this information the effective Hamiltonian that describes the system is given by

\begin{eqnarray} \label{ham}
H_{\mathrm{eff}}&=&(E_{3}+\frac{D}{2}M_{3}^{2})\sum_{M_{3}}|M_{3}\rangle \langle
M_{3}|+E_{2}\sum_{\alpha M_{2}}|\alpha M_{2}\rangle \langle \alpha M_{2}|\nonumber\\
&& + \sum_{\nu k\alpha \sigma }\epsilon _{\nu k}c_{\nu k\alpha \sigma
}^{\dagger }c_{\nu k\alpha \sigma }+\\
&&\sum_{\alpha\nu k\sigma }V_{\nu }\langle \tfrac{3}{2}
M_{3}  |1\tfrac{1}{2};M_{2}\sigma \rangle 
( | M_{3}\rangle \langle \alpha M_{2}|c_{\nu k\alpha \sigma }+\mathrm{H.c.} ),  \nonumber 
\end{eqnarray}
where $E_n$, $M_n$ represent the energies and the spin projections along the chain, which was chosen as 
the 
quantization axis, of states with $n=2, 3$ holes in the $3d$ shell of Co. The term 
$\langle SM| S_2 S_1; M_2 M_1 \rangle$ stands for the 
standard SU(2) Clebsch-Gordan coefficients.
The state with three holes
and maximum spin projection is denoted by $\vert 3/2 \rangle = 
\hat{d}_{xz\downarrow}^{\dagger}\hat{d}_{yz\downarrow}^{\dagger}\hat{d}_{xy\downarrow}^{\dagger}
\vert 0 \rangle$, where $\vert 0 \rangle$ represents the $3d^{10}$ configuration and the operators
$\hat{d}_{\beta\sigma}^{\dagger}$ create a hole with symmetry $\beta$ and spin projection $\sigma$.
The states with two holes in Co can be constructed by removing an $\alpha$ ($\alpha=xz, yz$) hole. 
For instance, the maximum spin projection of the state with two holes is given by
$\vert \bar{\alpha}, 1 \rangle = \hat{d}_{\alpha\downarrow}\vert 3/2 \rangle$. Then the state 
$\vert \bar{\alpha}, 1 \rangle$ does not contain the hole with symmetry $\alpha$, which jumped to the Au
band. In other words, if the electron $5d_{xz}$ of Au jumps to the Co, the remaining state 
$\vert xy, 1 \rangle$ has holes in $3d_{yz}$ and $3d_{xy}$, but not in $3d_{xz}$.
The other relevant states with two and three holes can be obtained by using the spin lowering operator.

There are also states with three holes in the Co atom coupled to a total spin $S=1/2$ in which one of the 
$\beta$-orbitals is doubly occupied but, in view of the obtained \textit{ab-initio} calculations, these 
can be considered as high exited states.

The operator $c_{\nu k\alpha \sigma}^{\dagger }$ creates a hole in the $5d$ shell of Au with symmetry 
$\alpha$ where $\nu=L, R$ denotes the left or the right side of Co atom, respectively.  The hopping $V_\nu$
defines the hybridization $\Gamma_\nu = 2\pi\sum_k \vert V_\nu \vert ^2 \delta(\omega-\epsilon _{\nu k})$,
neglecting the weak dependence with the energy $\omega$.
 
Consequently, the Hamiltonian in Eq. \ref{ham} describes fluctuations between the quadruplet 
of the $3d^7$ 
configuration and two triplets corresponding to the $3d^8$ configuration, with one $xy$-hole and the other in
either $\alpha=yz$ or $\alpha=xz$ symmetries, via the hybridization with 
the states of symmetry $\Delta_3$ of the Au leads. 

The last term in Eq. \ref{ham} represents the mixing Hamiltonian, $H_{mix}$, between the impurity and 
conducting Au atoms. Writing explicitly the non vanishing Clebsch-Gordan coefficients this becomes,

\begin{eqnarray} \label{h-mix}
H_{mix}&=&\sum_{\alpha\nu k\sigma }V_{\nu } 
   ( | \tfrac{3}{2} \rangle\langle\bar{\alpha}, 1 | +  
      \sqrt{\tfrac{2}{3}}| \tfrac{1}{2}\rangle\langle\bar{\alpha}, 0 | + \nonumber \\
     && \sqrt{\tfrac{1}{3}} |-\tfrac{1}{2}\rangle\langle\bar{\alpha},-1 | ) c_{\nu k\alpha \uparrow} \nonumber \\
     & +& 
   ( | -\tfrac{3}{2}\rangle\langle\bar{\alpha}, -1 | +  
      \sqrt{\tfrac{2}{3}}|-\tfrac{1}{2}\rangle\langle\bar{\alpha}, 0 | + \nonumber \\
     && \sqrt{\tfrac{1}{3}}| \tfrac{1}{2}\rangle\langle\bar{\alpha}, 1 | ) c_{\nu k\alpha \downarrow} + \mathrm{H.c.}.
\end{eqnarray}

The operator that creates a hole in the $3d$ shell of the Co atom with symmetry $\alpha$ and spin $\sigma$
can be represented by Hubburd operators between states containing two and three holes. In that case, the 
two holes forming the triplet are in the $xy$ and $\bar{\alpha}$ orbitals. It can be written as follows

\begin{eqnarray} \label{operador-fisico}
 d_{\alpha\uparrow}^{\dagger}&=& | \tfrac{3}{2} \rangle\langle\bar{\alpha}, 1 | +  
      \sqrt{\tfrac{2}{3}}| \tfrac{1}{2}\rangle\langle\bar{\alpha}, 0 | + 
      \sqrt{\tfrac{1}{3}}|-\tfrac{1}{2}\rangle\langle\bar{\alpha},-1 |, \\
 d_{\alpha\downarrow}^{\dagger}&=& | -\tfrac{3}{2}\rangle\langle\bar{\alpha}, -1 | +  
      \sqrt{\tfrac{2}{3}}|-\tfrac{1}{2}\rangle\langle\bar{\alpha}, 0 | + 
      \sqrt{\tfrac{1}{3}}| \tfrac{1}{2}\rangle\langle\bar{\alpha}, 1 | . \nonumber 
\end{eqnarray}

Then, the hybridization term of the Hamiltonian takes the usual Anderson impurity form, that is

\begin{eqnarray} \label{h-mix-2}
H_{mix}=\sum_{\alpha\nu k\sigma }V_{\nu }(  d_{\alpha\sigma}^{\dagger}c_{\nu k\alpha \sigma} + 
                                          c_{\nu k\alpha \sigma}^{\dagger}d_{\alpha\sigma}) .
\end{eqnarray}

Depending on the value of the anisotropy $D$, interesting and different physics emerges from
this Hamiltonian. The role played by $D$ is to split the quadruplet states into two doublets
with spin projection $M_3=\pm 3/2$ and those with $\pm 1/2$. 
For the case in which the anisotropy vanishes, $D=0$, 
in addition to the SU(2) channel symmetry  
the Hamiltonian also has the rotational spin SU(2) one. For this isotropic case, the model
in Eq. \ref{ham} reduces to the underscreened impurity Anderson model, in which the two channels
with spin $1/2$ compensate a part of the total impurity spin $3/2$. For one channel, the model
was solved exactly by Aligia \textit{et al.}, by using the Bethe ansatz 
\cite{aligia-balseiro-proeto} 
for the spin 1 underscreened one-channel Kondo model,  a singular FL ground state was found \cite{mehta}.
It is natural to expect
similar physics our case for $D=0$, which corresponds to a spin 3/2 underscreened by two conduction channels. 
The behavior of the conductance at low temperatures indicates
this is actually the case \cite{solange-1}.

When the anisotropy takes positive values, as it was found for the realistic case of the setup
in Fig. \ref{setup}b, $D=1.7$ meV, the doublet with $M_3=\pm 1/2$ spin projections becomes the 
one with lowest energy. Therefore, the two channels with spin $1/2$ overscreened the effective impurity 
spin $1/2$. Signatures of NFL behaviour in both, the impurity contribution to the entropy and the
equilibrium conductance, were previously reported \cite{solange-1}. In general, the behaviour 
of the model at low energies agrees with the corresponding one to the two-channel Kondo problem
 \cite{aff, emery}.

Finally, for negative values of the anisotropy $D$, which could be achieved by stretching the chain,
there is no Kondo physics. This follows from the fact that the doublet $M_3=\pm 3/2$ is now the 
fundamental one and the two channels with spin $1/2$ cannot flip the projections $\pm 3/2$ into 
each other. A residual entropy at zero temperature of $\ln(2)$ was found \cite{solange-1} in this case
and agrees with the non screened doublet  $M_3=\pm 3/2$ at the impurity site.

\section{Numerical Results}\label{numerical-results}

In this section we present an accurate solution of the local spectral function 
by using the numerical renormalization group (NRG) 
as well as 
the differential conductance at the Co site obtained within the non crossing approximation (NCA).

For the numerical calculations, the complete set of the parameters determining the model was 
extracted from first principle calculations and are reported in Ref. \onlinecite{solange-1}. Here
we summarize the parameters obtained. The total resonant level 
width $\Gamma = 0.6$ eV, is determined from the width of the peak of the degenerate $xz, yz$ 
states above the Fermi energy.  From the average position of these peaks, we define the charge transfer 
energy to be $E_{32}=E_{3}-E_{2}=-0.3$ eV and finally, we take the conduction $5d_{xz,yz}$ bands extending 
from $-W$ to $W$ with 
$W=5$.

\subsection{Impurity spectral density near the quantum critical point (NRG)} \label{rho-NRG}

The impurity spectral function at the impurity site per channel and spin is given by $
 \rho_{\alpha\sigma}(\omega) = -\frac{1}{\pi} G^r_{\alpha\sigma}(\omega) $
where $G^r_{\alpha\sigma}(\omega)$ is the Fourier transform of the retarded 
Green function also per channel and spin,
 
\begin{equation}
 G^r_{\alpha\sigma}(t) = -i\theta(t) \langle \{\hat{d}_{\alpha\sigma}(t),
\hat{d}^{\dagger}_{\alpha\sigma}(0)\} \rangle.
\end{equation}

The equilibrium conductance through the Co atom as a function of temperature, $G(T)$, 
depends on the total spectral function, 

\begin{equation}
 \rho(\omega) = \sum_{\alpha\sigma} \rho_{\alpha\sigma}(\omega),
\end{equation}
and it is given by

\begin{equation}\label{conductance}
 G(T) =  G_0 \frac{\pi\Delta A }{2}\int ~d\omega (-f'(\omega)) \rho(\omega),
\end{equation}
where $G_0=2e^2/h$ is the quantum of conductance, $f(\omega)$ is the Fermi function, 
$\Delta = \frac{\Gamma}{2} = \Delta_L + \Delta_R$ represents the total resonant level width, 
and $A=4\Delta _{L}\Delta _{R}/\Delta ^{2}$ 
stands for the asymmetric connection between the Co atom and the left and right leads.   

As it was previously mentioned, the model presents rich physics depending on the value of the 
anisotropy $D$. 
For positive $D$ and low temperatures, the conductance was successfully scaled by
$G(T)=a-b \sqrt{T}$, a behaviour similar to the two-channel Kondo model. Furthermore,
$a=G(0)$,  was found to be near $G_0$, neglecting the small asymmetry 
between the leads given by the factor $A=0.977$, see Fig. 4 of Ref. \onlinecite{solange-1}.
This is half the value expected for a Fermi liquid with two channels in the unitary limit (Kondo regime).
Consequently, within the overscreened regime, this seems to force the spectral density 
at the Fermi level (chosen to be at $\omega =0$),
to be $\rho(0)\sim\frac{2}{\pi\Delta}$ while the spectral weight per channel and spin seems to be
$\rho_{\alpha\sigma}(0)\sim\frac{1}{2\pi\Delta}$.  
This agrees with the known rule for the 2CA impurity model \cite{anders}, which can 
be obtained from our model in the limit $D \rightarrow + \infty $ and also for the 2CK model
\cite{mit}.
Several features in the Kondo resonance distinguish the 1CA
from the 2CA  spin-1/2 model \cite{cox-sawadowski}. 
Among others, 
i) the spectral weight for 2CA at $\omega =0$ in the Kondo regime, is reduced to 
nearly half the value of 1CA 
$\rho_{\sigma}(0)\sim\frac{1}{\pi\Delta}$, 
ii) the resonance in 2CA is pinned at the Fermi
level in contrast to the slight shift 
to positive energies found in the 1CA.

Within the underscreened regime, $D=0$, the conductance at temperature $T=0$ seems to approach 
to $2G_0$ which implies that the total spectral weight at the Fermi level should be near 
$\rho(0)\sim\frac{4}{\pi\Delta}$. 
According to that, the spectral density at $\omega=0$ per 
channel and spin surprisingly agrees with the specified by the generalized Friedel sum rule,\cite{yoshi2}
for two orbitals assuming an impurity occupation near 1 for each of them: 
$\rho_{\alpha\sigma}(0)\sim\frac{1}{\pi\Delta}$.
This result differs from that of the well studied 1C underscreened Kondo effect in 
spin-1 molecules \cite{logan, cornaglia}, where it was found that the phase shift has 
an additional term $\pi/2$ to that due to the contribution of the displaced electrons \cite{logan}.

For $D<0$, the projections $M_3 = \pm 3/2$ are not connected by the hopping processes with both channels. 
One has not a spin Kondo effect and there is no rule for the spectral density at the Fermi level.
It is expected a continuous reduction of that weight when increasing $\vert D\vert$ as a consequence 
of the vanishing Kondo resonance. In the limit of $ D\longrightarrow - \infty$ the model splits into two 
different resonant models and Fermi liquid results could be used.

Important questions arise at this point: 
How does the spectral density evolve near to the transition point from negative to positive values 
of the anisotropy D ? Does the transition constitute a crossover or a quantum critical phase transition ?
Are there more energies scales involved in the transition in addition to the well known Kondo 
scale?

In what follows, we present the results of the spectral density
from NRG calculations for different values of the anisotropy $D$. 
The results are obtained from full density matrix (fdm-)
NRG calculations which exploited the SU(2) channel symmetry
together with the abelian U(1) symmetries for total
spin and total charge.\cite{andreas-1, andreas-2, andreas-3}
Further NRG specific parameters 
are $\Lambda=4$ for the logarithmic discretization
of the conduction bands together with $z$-averaging using
$N_z=2$, \cite{zrefs} and a truncation energy $E_{tr} = 10$
in rescaled units (as defined in [\onlinecite{andreas-3}]). This
resulted in retaining up to 16\,000 multiplets (53\,000 states)
per iteration with exact diagonalization of state spaces of
dimension up to 234\,000 multiplets (846\,000 states).
The estimated resulting
discarded weight of $\delta\rho < 10^{-15}$
indicates numerically well-converged data.\cite{andreas-4}

We start by defining the Kondo temperature, $T_K$, for $D=0$ as the half width 
at the half maximum of the spectral density. For the model parameters representing 
the trimer it was found to be $T_K \approx 7\cdot10^{-6}$ eV.

\begin{figure}[tbp]
\includegraphics[clip,width=7cm]{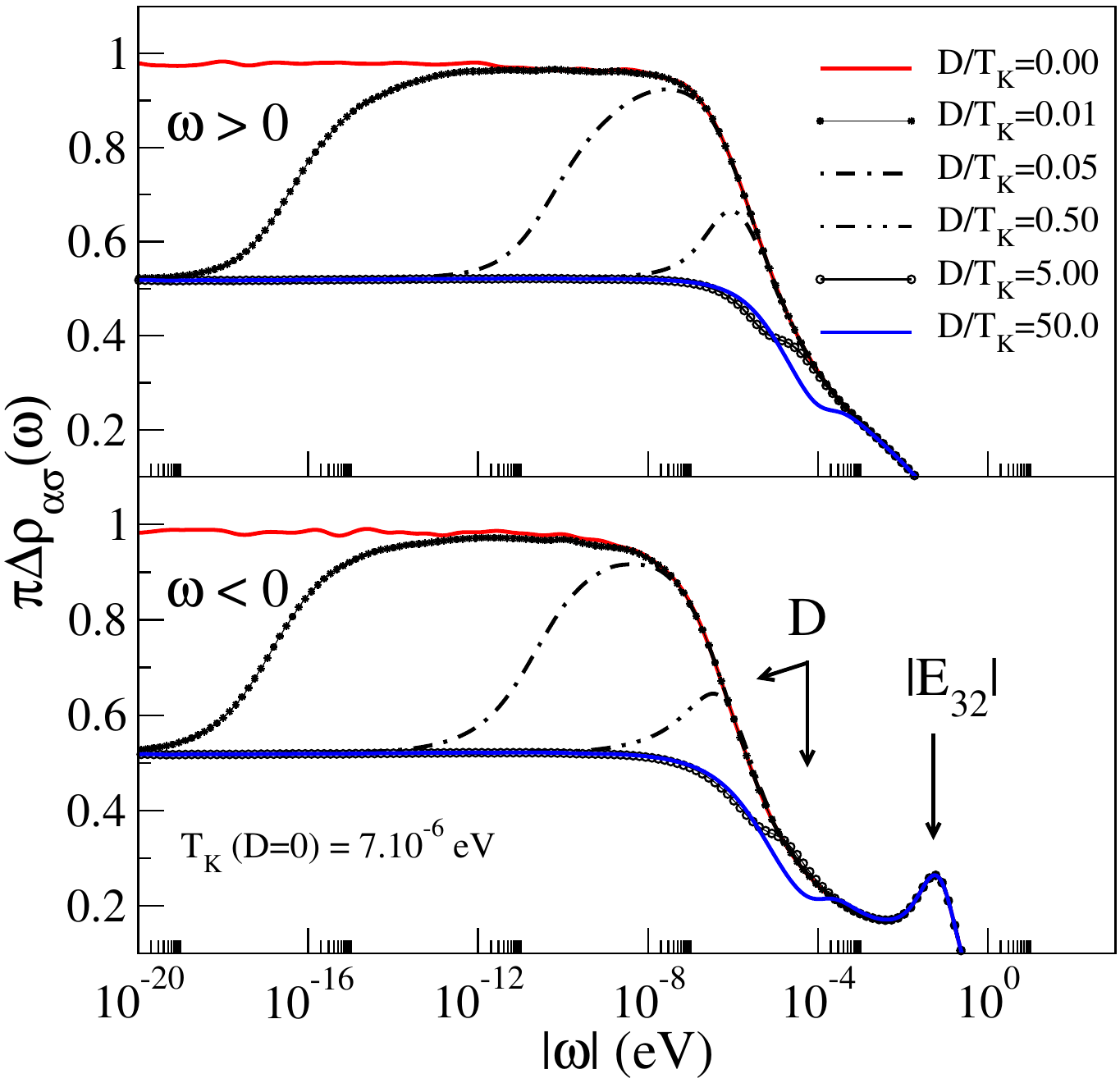}
\caption{(color online). Spectral density of the Co atom as a function of frequency in a 
logarithmic scale for different values of $D$ (NRG).
The upper and lower panels show the 
data for positive and negative $\omega$ respectively. The arrows stand for the charge transfer energy $E_{32}$ and the 
anisotropy $D$.}
\label{rho-1}
\end{figure}

\begin{figure}[tbp]
\includegraphics[width=1\linewidth]{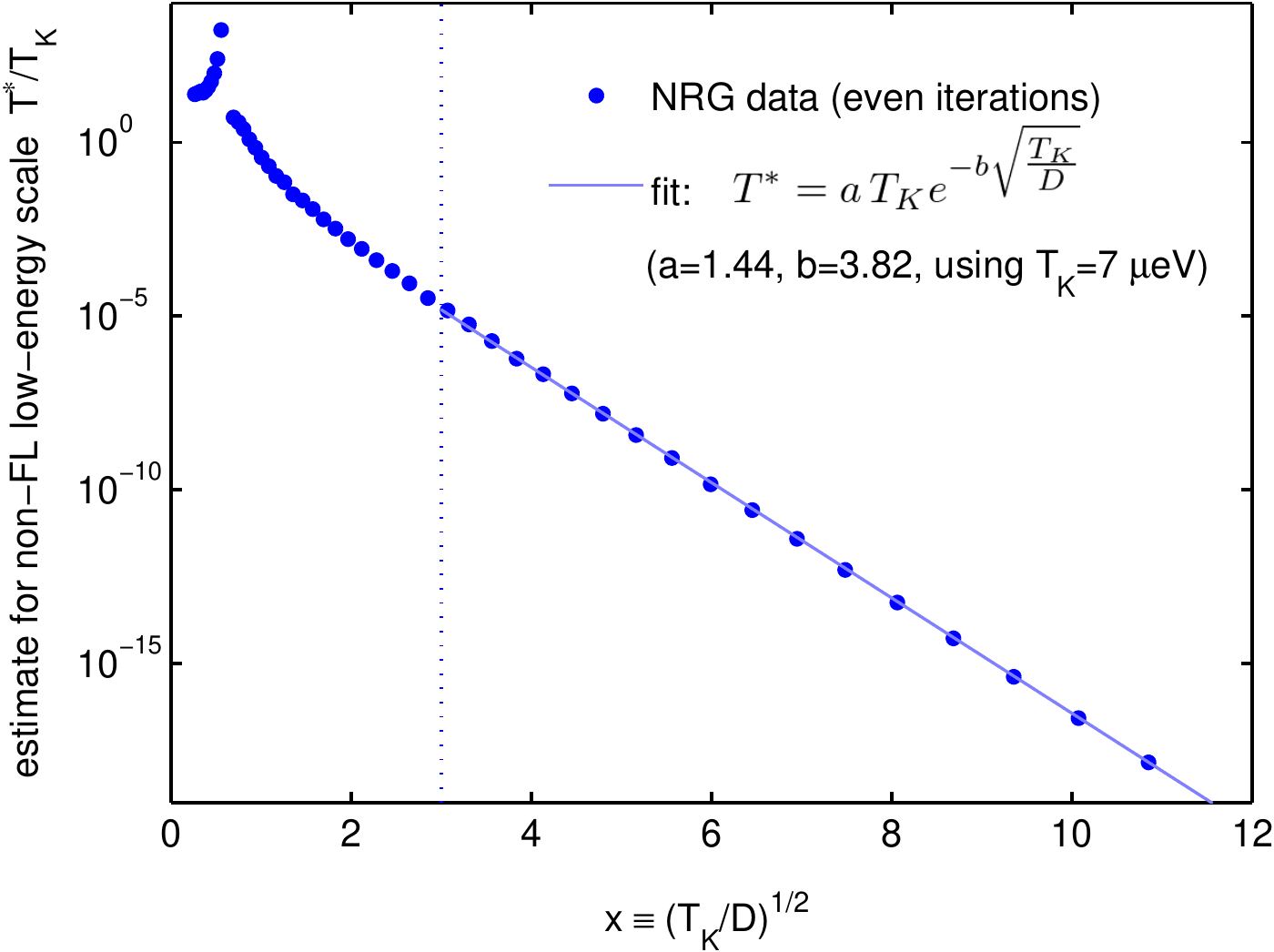}
\caption{(color online). Analysis of the low-energy 
scale $T^\ast$ extracted from the NRG data (using even
iterations $n$, with similar data and conclusions for
odd iterations [not shown]).
For the determination of $T^\ast$, the exponential
convergence of the rescaled and iteratively subtracted ground
state energy $E_0^{(n)}$ at iteration $n$ deep into the low-energy
fixed point was extrapolated towards larger energies
(smaller $n$), such that $\Delta \tilde{E}_0^{(n^\ast)} := 1$
(in rescaled energies) with $\Delta \tilde{E}_0^{(n)}$
the plain exponential fit to $|E_0^{(n)} - E_0^{(\infty)}| < 0.01$.
With this, the low energy scale $T^\ast := \Lambda^{-n^\ast/2}$
was determined individually for each value of the anisotropy $D$.
Finally, the fit to the analytical expression for $T^*$
was carried out for $x>3$ (indicated by the vertical dotted
line), i.e. $D<0.11~T_K$.}
\label{fig:LTscale}
\end{figure}

In Fig. \ref{rho-1} we show the impurity spectral density per channel and spin for 
several positive values of the anisotropy $D$ in units of $T_K$.
By analogy with the Anderson model, the spectral data is scaled by $\pi\Delta$. In addition to the 
peak at the charge transfer energy at $\omega=E_{32}$, the spectral density exhibits two shoulders or satellite 
peaks at energies related with the anisotropy at $\omega=\pm D$
for $D >T_K$ . These energies are indicated with arrows in the 
lower panel, corresponding to negative frequencies.
Within the underscreened regime, as it was previously mentioned, the 
maximum of the spectral density at low energy
roughly agrees with the expected one for FL behaviour.
From the Friedel sum rule, one would expect $\pi\Delta\rho_{\alpha\sigma}(0)=\sin^2(\pi \langle n_{\alpha 
\sigma}\rangle)$ \cite{yoshi2}
that is near to the result reported in Fig. 2 for an occupation of the Co atom per channel and spin
$\langle n_{\alpha \sigma}\rangle \simeq 1/2$. Using the real parameters representing the system shown in 
Fig. \ref{setup}b, we obtain $\langle n_{\alpha \sigma}\rangle = 0.428$. 
Therefore, some degree 
of intermediate valence is present which would lead to a 5\% lower value, i.e. $\pi\Delta\rho_{\alpha\sigma}(0) \simeq 
0.95$.
To the best of our knowledge, however, there are no exact results for the spectral weight at $\omega=0$ for this 
kind of model, because it is expected to be a singular FL, in which the Friedel sum rule is not valid. 
In the case of the 1C underscreened spin $S=1$, D. Logan \textit{et al.} have found, on the 
basis of NRG calculations, that when the total
impurity occupation, $ \langle n_{imp} \rangle$, tends to 2, the normalized spectral weight at the Fermi level, 
$\pi\Delta\rho_{\alpha\sigma}(0)$, approaches  1 instead of zero as expected for a FL \cite{logan}. 

When the anisotropy $D$ goes from $D=0$ to $D>0$, the spectral weight at the Fermi level
is reduced suddenly to half its value.
As it was previously mentioned, this agrees with the results expected for the 
2CA model. When $D$ turns positive, there is a new low-energy scale entering the system which controls the 
crossover from the underscreened to overscreened phases.
An analysis of the low-energy spectrum
as shown in Fig.~\ref{fig:LTscale}, clearly suggests for $D\ll T_K$ 
the asymptotic form 

\begin{equation}
 T^{\ast} = a T_{K}~e^{ -b(T_{K}/D)^{1/2}},
\end{equation}

with $a$ and $b$ some dimensionless constants of order 1. 
It is interesting to note that 
in the case of 1C underscreened spin-1 
model including anisotropy, a similar scale was found \cite{cornaglia}. 

\begin{figure}[tbp]
\includegraphics[clip,width=7cm]{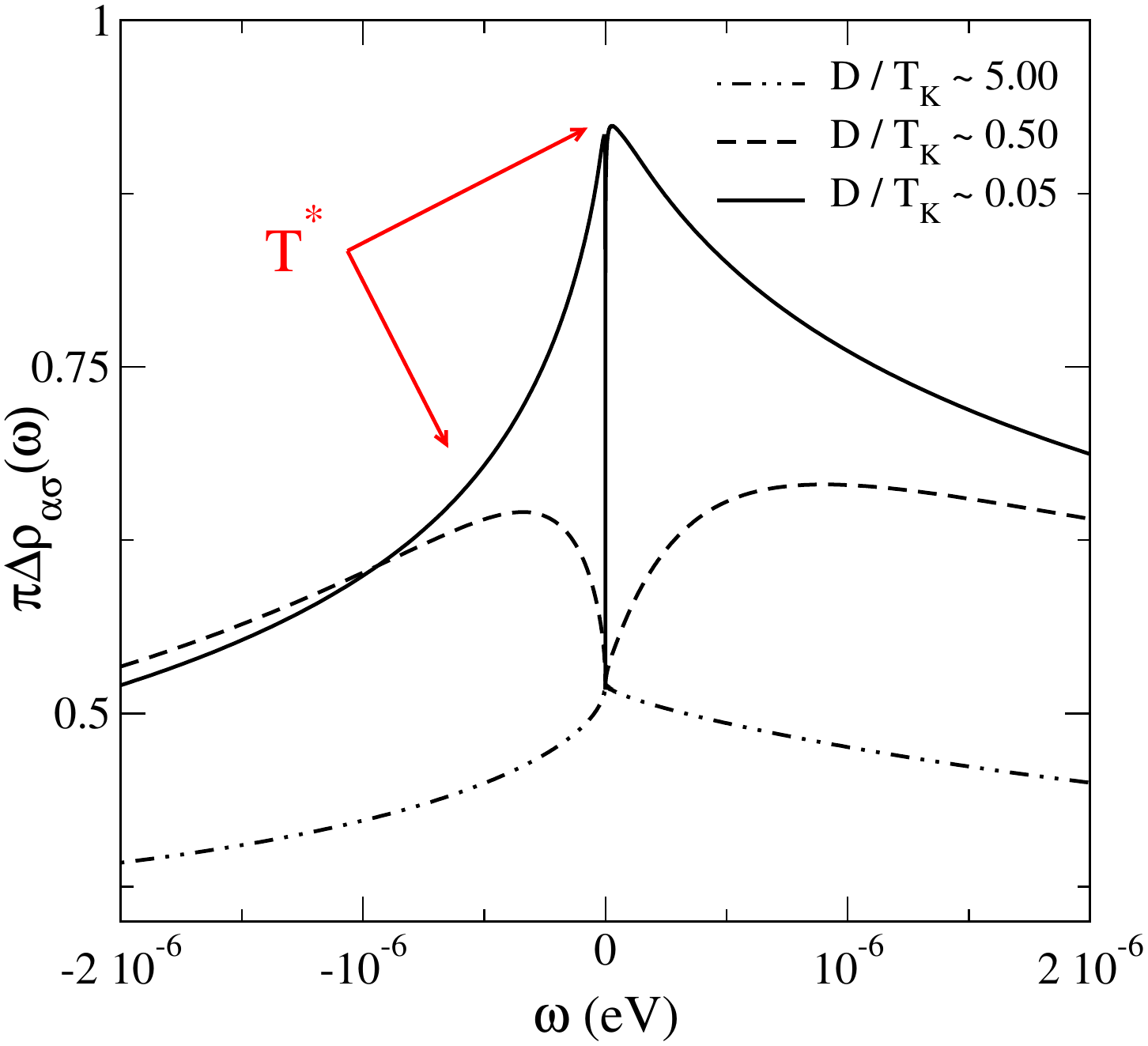}
\caption{(color online). Spectral density per channel and spin for different values of the ratio $D/T_K$ (NRG).}
\label{rho-2}
\end{figure}

As it can be seen in Fig. \ref{rho-2}, for $D \gg T_K$, 
the spectral weight continuously increases when the energy 
approachs to zero.
However, for $D<T_K$ the spectral weight increases until $\vert \omega\vert \sim T^{\ast}$.
For $\vert \omega\vert < T^{\ast}$ a dip is opened and the spectral density $\rho_{\alpha\sigma}(0)$ 
is suppressed to nearly half its value for $D=0$. In Fig. \ref{rho-2} we show this peculiar 
behaviour for two selected values of the ratio, $D/T_K = 0.5$ and  $D/T_K = 0.05$.

The results presented in Fig. \ref{rho-1} and Fig. \ref{rho-2} demonstrate that a QCP 
separating the overscreened and underscreened phases is present and 
NFL properties are obtained for any $D>0$.

\begin{figure}[tbp]
\includegraphics[clip,width=7cm]{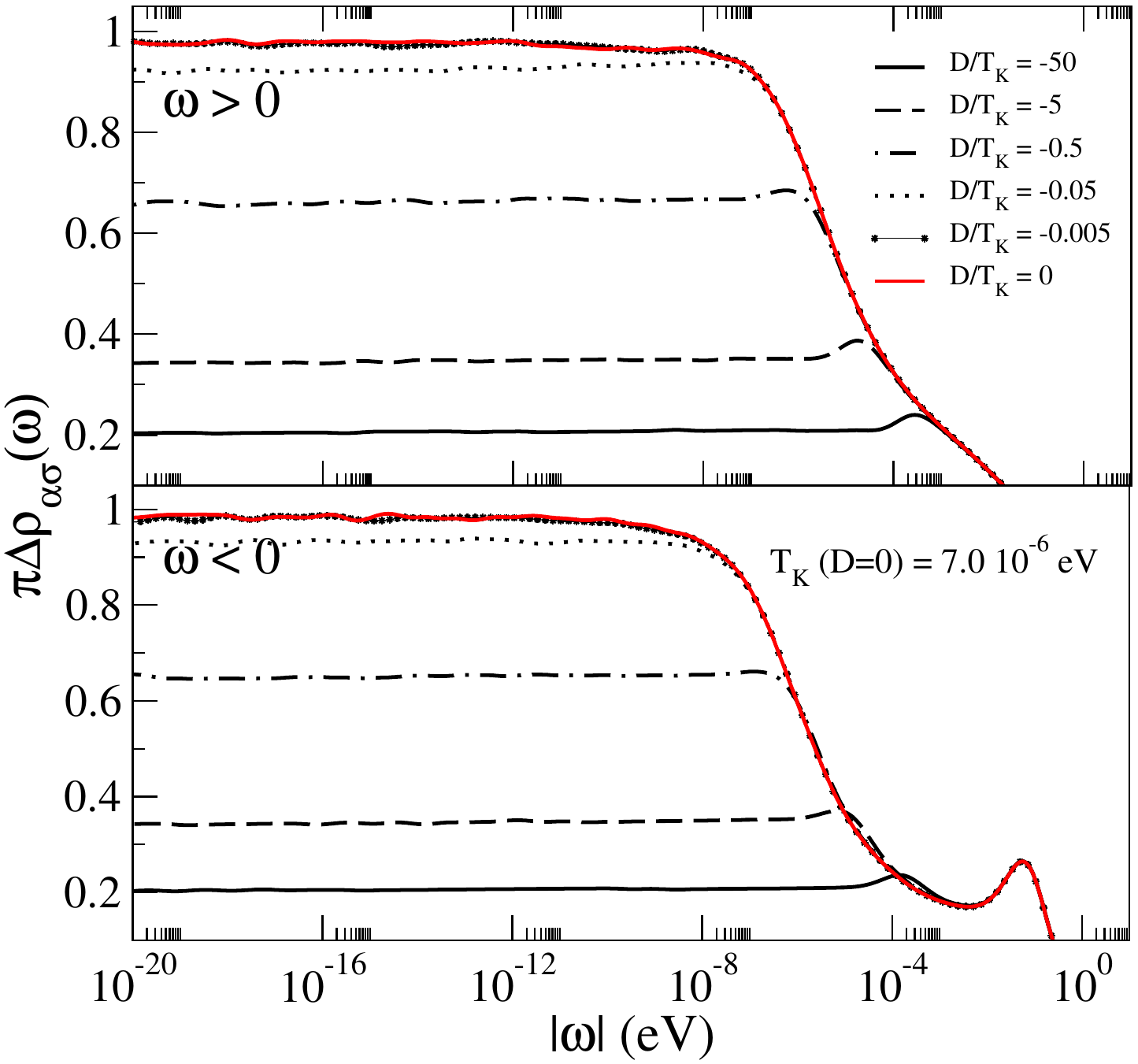}
\caption{(color online). Spectral density per channel and spin for different negative values of the ratio 
$D/T_K$ (NRG).}
\label{rho-3}
\end{figure}

Finally, in Fig. \ref{rho-3} we present the results for negative values of the anisotropy $D$.
As it can be seen, the spectral weight at the Fermi energy is continuously reduced when 
the values of $\vert D \vert$ are increased. This is expected due to the vanishing Kondo effect 
and follows from the fact that there is no spin-flip processes connecting the $M_3=\pm 3/2$ 
projections.

\subsection{Non equilibrium transport properties (NCA)}

In this section we analyse the differential conductance, $G(T,V)$, through the Co atom when a finite bias 
voltage is applied to the system. For this purpose we calculated the current through the Co atom for 
each temperature as a 
function of bias voltage, $I(V)$, being $G = dI / dV$. The current per channel can be expressed in 
terms of the corresponding spectral density \cite{meir} obtained in the presence of the two different 
chemical potentials $\mu_L$ and $\mu_R$, 

\begin{equation}
 I_{\alpha}(V) = \frac{2\pi e}{h} \frac{\Gamma_L\Gamma_R}{\Gamma_L+\Gamma_R}\sum_{\sigma} 
 \int d\omega~[f_L(\omega)-f_R(\omega)]~\rho_{\alpha\sigma}(\omega).
\end{equation}

While for the equilibrium situation we obtain the spectral density exactly by using the NRG, within 
the non equilibrium one we employ the non crossing approximation, NCA, to solve the model. 
To apply the NCA we introduce auxiliary bosons for the triplets ( $t_{\alpha M_2}$) and 
auxiliary fermions ($q_{M_{3}}$) for the quadruplet. In terms of the auxiliary
operators the Hamiltonian in Eq.(\ref{ham}) takes the form

\begin{eqnarray}\label{ham-aux-particles}
H_{\mathrm{eff}}& = &(E_{3}+\frac{D}{2}M_{3}^{2})\sum_{M_{3}} q^{\dagger}_{M_{3}}q_{M_{3}}
+E_{2}\sum_{\alpha M_{2}} t^{\dagger}_{\alpha M_2}t_{\alpha M_2} + \nonumber\\
&&\sum_{\nu k\alpha \sigma }\epsilon _{\nu k}c_{\nu k\alpha \sigma
}^{\dagger }c_{\nu k\alpha \sigma }+\\
&&\sum_{\alpha\nu k\sigma }V_{\nu }\langle \tfrac{3}{2}
M_{3}  |1\tfrac{1}{2};M_{2}\sigma \rangle 
( q^{\dagger}_{M_{3}}t_{\bar{\alpha} M_2}c_{\nu k\alpha \sigma }+\mathrm{H.c.} ).  \nonumber 
\end{eqnarray}

The expression for the physical operators $d_{\alpha\sigma}^{\dagger}$ in terms of auxiliary particles 
are given by

\begin{eqnarray} \label{operador-fisico-aux-part}
 d_{\alpha\uparrow}^{\dagger}&=& q^{\dagger}_{3/2}t_{\bar{\alpha},+1} +  
      \sqrt{\tfrac{2}{3}}q^{\dagger}_{1/2}t_{\bar{\alpha},0}  + 
      \sqrt{\tfrac{1}{3}}q^{\dagger}_{-1/2}t_{\bar{\alpha},-1}, \\
 d_{\alpha\downarrow}^{\dagger}&=& q^{\dagger}_{-3/2}t_{\bar{\alpha},-1} +  
      \sqrt{\tfrac{2}{3}}q^{\dagger}_{-1/2}t_{\bar{\alpha},0}  + 
      \sqrt{\tfrac{1}{3}}q^{\dagger}_{1/2}t_{\bar{\alpha},+1}.    
\end{eqnarray}

Since only one state should be occupied at each time, the total operator number of auxiliary 
particles must satisfy the following constraint
 
\begin{equation}
 \sum_{M_{3}} q^{\dagger}_{M_{3}}q_{M_{3}} + \sum_{\alpha M_{2}} t^{\dagger}_{\alpha M_2}t_{\alpha M_2} = 1.
\end{equation}

The spectral density associated with the operator $d_{\alpha\sigma}^{\dagger}$ can be obtained by 
convolution of the greater and lesser Green functions of the auxiliary particles. For instance, 
the $\rho_{\alpha\uparrow}(\omega)$ is given by

\begin{eqnarray}
 \rho_{\alpha\uparrow}(\omega) = \frac{-1}{4\pi^2 Q} \int d\omega' \{ 
 & ( & G^{>}_{3/2}(\omega+\omega')G^{<}_{\alpha,+1}(\omega')- \nonumber \\
 &&      G^{<}_{3/2}(\omega+\omega')G^{>}_{\alpha,+1}(\omega')) + \nonumber  \\
 & \tfrac{2}{3} ( & G^{>}_{1/2}(\omega+\omega')G^{<}_{\alpha,0}(\omega')-\\
 &&  G^{<}_{1/2}(\omega+\omega')G^{>}_{\alpha,0}(\omega') ) + \nonumber  \\
 & \tfrac{1}{3} ( & G^{>}_{-1/2}(\omega+\omega')G^{<}_{\alpha,-1}(\omega')- \nonumber \\
 &&  G^{<}_{-1/2}(\omega+\omega')G^{>}_{\alpha,-1}(\omega') ) \}, \nonumber 
\end{eqnarray}

being $Q$ the impurity canonical partition function
\begin{equation}\label{Q-number}
 Q =  \frac{-i}{2\pi} \int d\omega~ (~ \sum_{M_3} G^{<}_{M_3}(\omega) -
      \sum_{\alpha, M_2} G^{<}_{M_2}(\omega)~ ).
\end{equation}
A similar expression allows to obtain $\rho_{\alpha\downarrow}(\omega)$ and, in absence of
an applied magnetic field, as it is actually our case, 
$\rho_{\alpha\downarrow}(\omega)=\rho_{\alpha\uparrow}(\omega)$.

In order to obtain the greater auxiliary Green function, a selfconsistent loop for
the selfenergies have to be solved,

\begin{eqnarray}\label{lesser-system}
\Sigma_{q_{3/2}}^{>}(\omega) &=& \frac{1}{2\pi}\sum_{\nu\alpha}\Gamma_{\nu\alpha\uparrow} \nonumber
              \int d\omega'  f(\omega'-\omega+\nu_\mu) G^{>}_{\bar{\alpha},+1}(\omega')  \nonumber \\
\Sigma_{q_{-3/2}}^{>}(\omega) &=& \frac{1}{2\pi}\sum_{\nu\alpha}\Gamma_{\nu\alpha\downarrow} \nonumber
              \int d\omega'  f(\omega'-\omega+\nu_\mu) G^{>}_{\bar{\alpha},-1}(\omega')   \\ \nonumber
\Sigma_{q_{1/2}}^{>}(\omega) &=& \frac{1}{6\pi}\sum_{\nu\alpha} \nonumber
              \int d\omega' f(\omega'-\omega+\nu_\mu) [~ \nonumber
              2\Gamma_{\nu\alpha\uparrow}G^{>}_{\bar{\alpha},0}(\omega')  \\ & \nonumber
              &+ \Gamma_{\nu\alpha\downarrow}G^{>}_{\bar{\alpha},1}(\omega')~] \nonumber \\ 
\Sigma_{q_{-1/2}}^{>}(\omega) &=& \frac{1}{6\pi}\sum_{\nu\alpha} \nonumber
              \int d\omega' f(\omega'-\omega+\nu_\mu) [~ \nonumber
              2\Gamma_{\nu\alpha\downarrow}G^{>}_{\bar{\alpha},0}(\omega')  \nonumber \\ &
              &+ \Gamma_{\nu\alpha\uparrow}G^{>}_{\bar{\alpha},-1}(\omega')~]  
\end{eqnarray}
together with the non-equilibrium Dyson equations

\begin{equation}\label{relacion-constitutiva}
 G_{i}^{\gtrless}(\omega) = G_{i}^{r}(\omega)\Sigma_{i}^{\gtrless}(\omega)G_{i}^{a}(\omega),
\end{equation}
where the retarded Green functions are given by 

\begin{equation}
 G_{i}^{r}(\omega) = \frac{1}{\omega-\epsilon_i-\Sigma_{i}^{r}(\omega)}.
\end{equation}
Within the NCA, the retarded and greater selfenergies are related by 
\begin{equation}
\Sigma_{i}^{>}(\omega) = 2i~ Im \Sigma_{i}^{r}(\omega).  
\end{equation}
An independent loop for the lesser selfenergies is needed to get the partition function and the 
lesser Green functions. The loop is closed using again the equations Eq.(\ref{relacion-constitutiva}),

\begin{eqnarray}\label{greater-system}
\Sigma_{q_{3/2}}^{<}(\omega) &=& -\frac{1}{2\pi}\sum_{\nu\alpha}\Gamma_{\nu\alpha\uparrow} \nonumber
              \int d\omega'  f(\omega-\omega'-\nu_\mu) G^{<}_{\bar{\alpha},+1}(\omega')  \nonumber \\
\Sigma_{q_{-3/2}}^{<}(\omega) &=& -\frac{1}{2\pi}\sum_{\nu\alpha}\Gamma_{\nu\alpha\downarrow} \nonumber
              \int d\omega'  f(\omega-\omega'-\nu_\mu) G^{<}_{\bar{\alpha},-1}(\omega')  \nonumber \\
\Sigma_{q_{1/2}}^{<}(\omega) &=& -\frac{1}{6\pi}\sum_{\nu\alpha}\nonumber
              \int d\omega' f(\omega-\omega'-\nu_\mu) [~ \nonumber
              2\Gamma_{\nu\alpha\uparrow}G^{<}_{\bar{\alpha},0}(\omega')  \\ \nonumber
              && + \Gamma_{\nu\alpha\downarrow}G^{<}_{\bar{\alpha},1}(\omega')~] \nonumber \\ 
\Sigma_{q_{-1/2}}^{<}(\omega) &=& -\frac{1}{6\pi}\sum_{\nu\alpha}  \nonumber
              \int d\omega' f(\omega-\omega'-\nu_\mu) [~\nonumber
              2\Gamma_{\nu\alpha\downarrow}G^{<}_{\bar{\alpha},0}(\omega') \nonumber \\
              &&+ \Gamma_{\nu\alpha\uparrow}G^{<}_{\bar{\alpha},-1}(\omega')~].  
\end{eqnarray}

The equilibrium properties can be simply obtained by setting $\mu_L = \mu_R = 0$ when solving the 
selfconsistent loop for the lesser and greater selfenergies. 

In the numerical procedure to solve the previous NCA equations, we follow the computational
algorithms, that ensure an accurate solution of the problem, detailed in Refs. \onlinecite{wingreen, hettler, roura}.

In what follows we present the NCA results for the impurity entropy, spectral density and differential 
conductance for different values of the anisotropy $D$.  

\begin{figure}[tbp]
\includegraphics[clip,width=7cm]{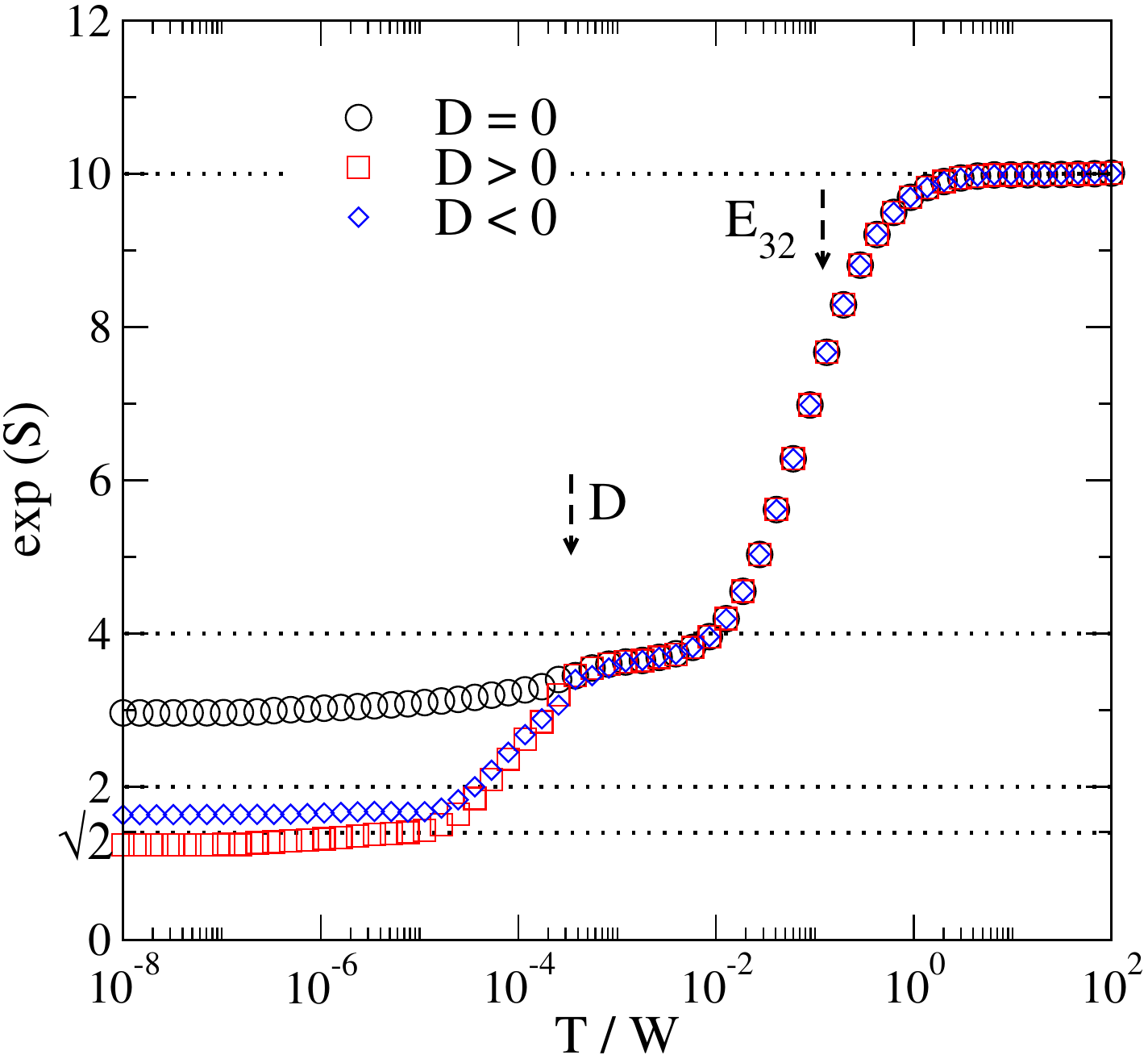}
\caption{(color online). NCA impurity contribution to the entropy (at equlibrium) as a function of temperature for 
$D=0$ and 
 $D=\pm0.0017$eV. Energies are given in units of the half bandwidth $W$.}
\label{entropia}
\end{figure}

In Fig. \ref{entropia} we show the Co contribution 
to the entropy as a function of temperature for different values of the anisotropy $D$. The latter
is calculated through numerical differentiation of the free energy \cite{hettler}.
At high temperatures all the impurity degrees of freedom are active due to the charge fluctuations and therefore 
the entropy tends to $S= \ln(10)$ (in units of the Boltzmann constant 
$k_B$). As the temperature is lowered, the charge transfer is frozen and the only active degrees of 
freedom correspond to the local moment regime characterized by a local spin 3/2, and therefore a plateau appears 
with $S \approx$ ln(4).  When the temperature reaches $T\sim\vert D \vert$, the three different regimes are 
separated. 
At low enough temperatures, our model is expected to have entropy $S= \ln(2)$  when $D<0$ reflecting the presence 
of two decoupled local moments. The same low temperature limit should be reached for the underscreened, $D=0$, 
regime in which a doublet is still present (see Fig. 3 of Ref. \onlinecite{solange-1}). As it can be seen from  
Fig. \ref{entropia}, the NCA overestimates the expected value for the underscreened case. This is related to 
the neglected vertex correction within this approach. 
Surprisingly, the NCA entropy at low temperatures 
for negative $D$ is closer to the expected one. Within the overscreened regime, 
the NCA entropy at low temperatures gives the correct residual value of 
$S\sim\frac{1}{2}\ln(2)$. It is well known that NCA is a reliable technique for the overscreened 2CA
model. Regarding the thermodynamic properties, the residual entropy and the scaling behaviour of the static 
magnetic susceptibility, among others, have been successfully compared with exacts Bethe ansatz 
results \cite{cox-kim}. 

\begin{figure}[tbp]
\includegraphics[clip,width=7cm]{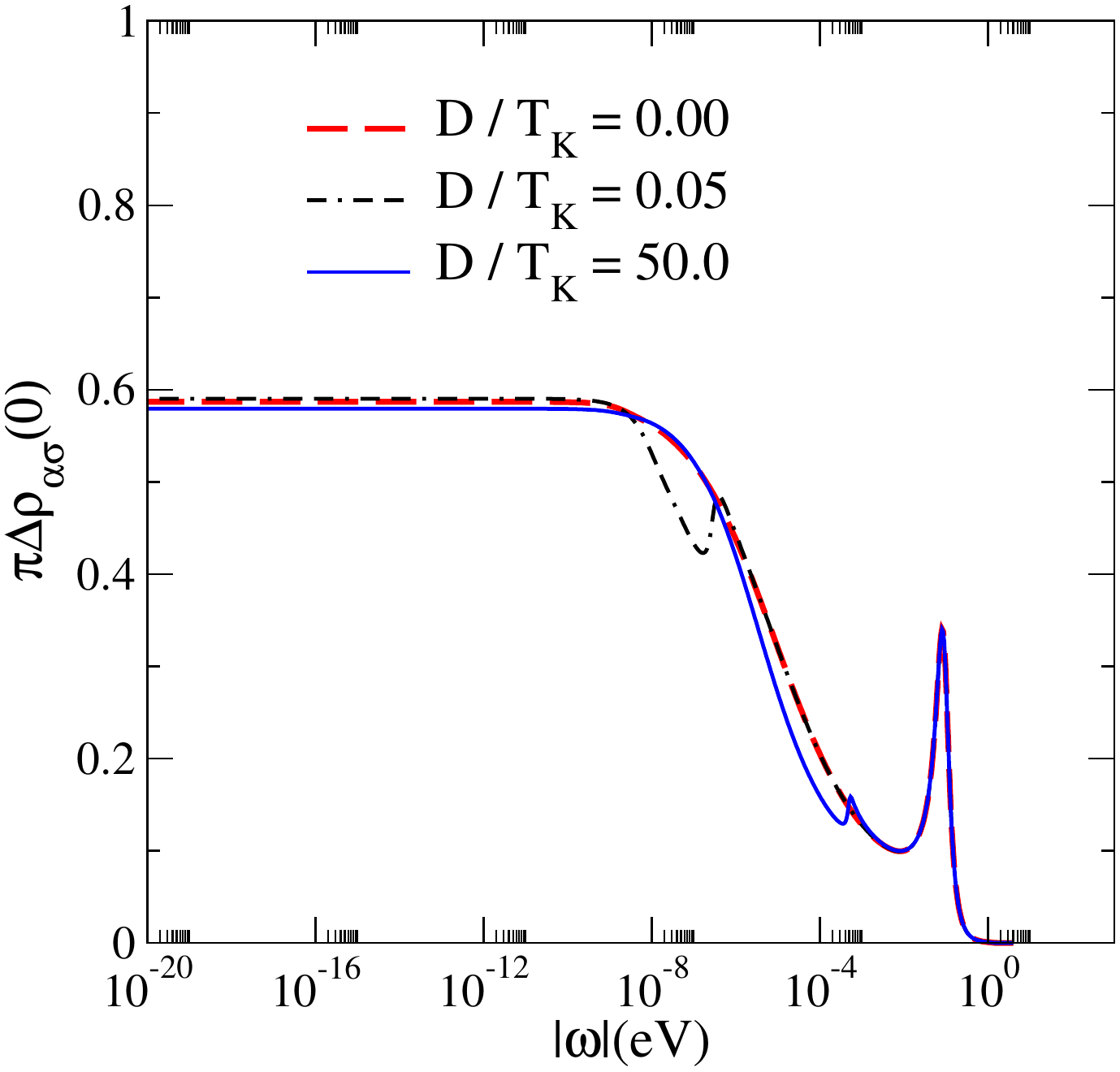}
\caption{(color online). NCA spectral density per channel and spin for different values of the ratio $D/T_K$ 
at equilibrium.}
\label{rho-nca}
\end{figure}

In Fig. \ref{rho-nca} we show the NCA results for the spectral density at equilibrium conditions for the same 
set of parameters as in Fig. \ref{rho-1} for several values of $D$. For simplicity, we show only the negative
frequency data.
When comparing the NCA results with the NRG corresponding ones, it is clear that for $D=0$, the NCA spectral 
weight at the Fermi level is strongly underestimated.
This can be understood as follows.
Within the underscreened regime, $D=0$, a scaling in the hybridization 
$\Delta'=\Delta/3$ in 
the system of NCA selfconsistent equations Eq.(\ref{lesser-system}), Eq.(\ref{greater-system}) 
and Eq.(\ref{Q-number}) leads to an identical system in which the ground and excited states have 
degeneracies $N=4$ and $M=6$, respectively. For such a model, the NCA spectral density at the Fermi level
is expected to be $\rho(0)\sim\frac{2\pi}{(N+M)^2 \Delta'}$, (see appendix B of Ref. 
\onlinecite{cox-kim}),\cite{factor-2}. 
We have verified that our calculations satisfy this rule. In addition, when $D$ becomes positive but lower
than the Kondo temperature associated with the underscreened case, the low-energy scale $T^\ast$ is completely 
absent. 
On the other hand, for large enough negative values of the anisotropy (not shown) no Kondo resonance is expected. 
However 
the NCA spectral function develop a spurious
spike at the Fermi level 
at low temperatures, in analogy with other cases of systems with a non degenerate 
ground state in abscence of hybridization 
(see Fig. 3 of Ref. \onlinecite{roura}). 
Therefore, we conclude 
from the comparison with NRG that the NCA approach does not represent a suitable technique for a 
quantitative and even qualitative treatment of the problem when $D < T_K$. 
 
Regarding the spectral function for $D \ge T_K$, we found the expected asymptotic low energy
dependence $\sim \sqrt{\omega}$ in the limit $\omega\longrightarrow 0$ and a slight overestimation of
the spectral weight at the Fermi level at very low temperatures. 
On the other hand, for dynamical properties the NCA reproduces the exact power law 
at low energies of all 4-point auxiliary correlation functions, like is the case of the spectral density 
\cite{cox-sawadowski, hettler}. Specifically, for transport properties like equilibrium conductance and 
differential conductance, both depending on the spectral density, the NCA gives the exact non trivial 
$\sqrt{T}$ and  $\sqrt{V}$ dependence respectively \cite{hettler-prl, solange-3}.

\begin{figure}[tbp]
\includegraphics[clip,width=7cm]{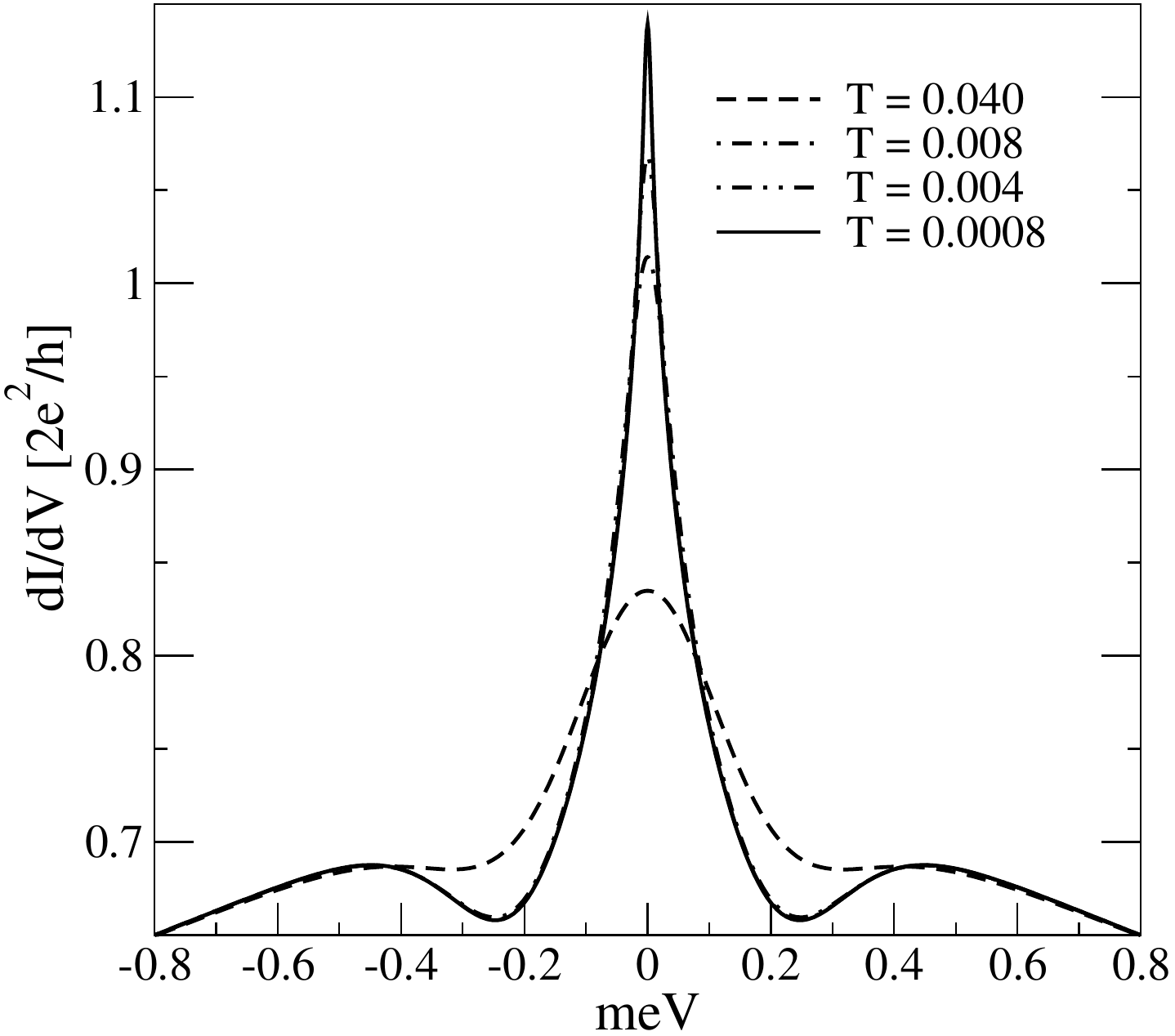}
\caption{Total differential conductance as a function of the bias voltage 
for several values of temperature.
The model parameters correspond to the real configuration of the trimer sketched 
in Fig.\ref{setup}(b) with the parameters given in the text.
Temperatures are given in units of meV. }
\label{didv}
\end{figure}

In Fig. \ref{didv} we present the total differential conductance through the Co atom for the real parameters
in the trimer geometry as a function of the bias voltage $V$ and for several temperatures. While 
for $T=V=0$, 
a value of 
$2G_0$ is expected in the general case of two independent conductance channels,
we obtain nealy half of it in agreement with NRG.
This is, once again, an additional verification of the non trivial 2C behaviour. The zero-bias anomaly 
(ZBA) reflects the low-lying energy dependence of the spectral function and a $\sqrt{V}$ behavior is explicitly
shown in the next two figures. A slight overestimation of the unitary limit is also obtained and follows
from the corresponding one of the spectral weight at the Fermi level.
In addition to the ZBA, the differential conductance exhibits two broad peaks located at 
$eV\sim \pm 2D$, related with the exited states $\pm 3/2$.
  
Affleck and Ludwig's conformal field theory (CFT) solution of the 2CK problem suggested 
that a scaling of the differential conductance $G(T,V)$ as a function of $T$ and $V$ should 
be possible with the form

\begin{equation}
 G(T,V) - G(T,0) = BT^{1/2} H(A \frac{eV}{k_B T}), 
\end{equation}
where H is an universal function with the conditions $H(0)=0$ and $H(x)\sim x^{1/2}$ for $x\gg1$,
and $A, B$ are non-universal constants (i.e. sample-dependent). The constant $B$ is determined from
the equilibrium conductance \cite{constant-B}.
  
\begin{figure}[tbp]
\includegraphics[clip,width=7cm]{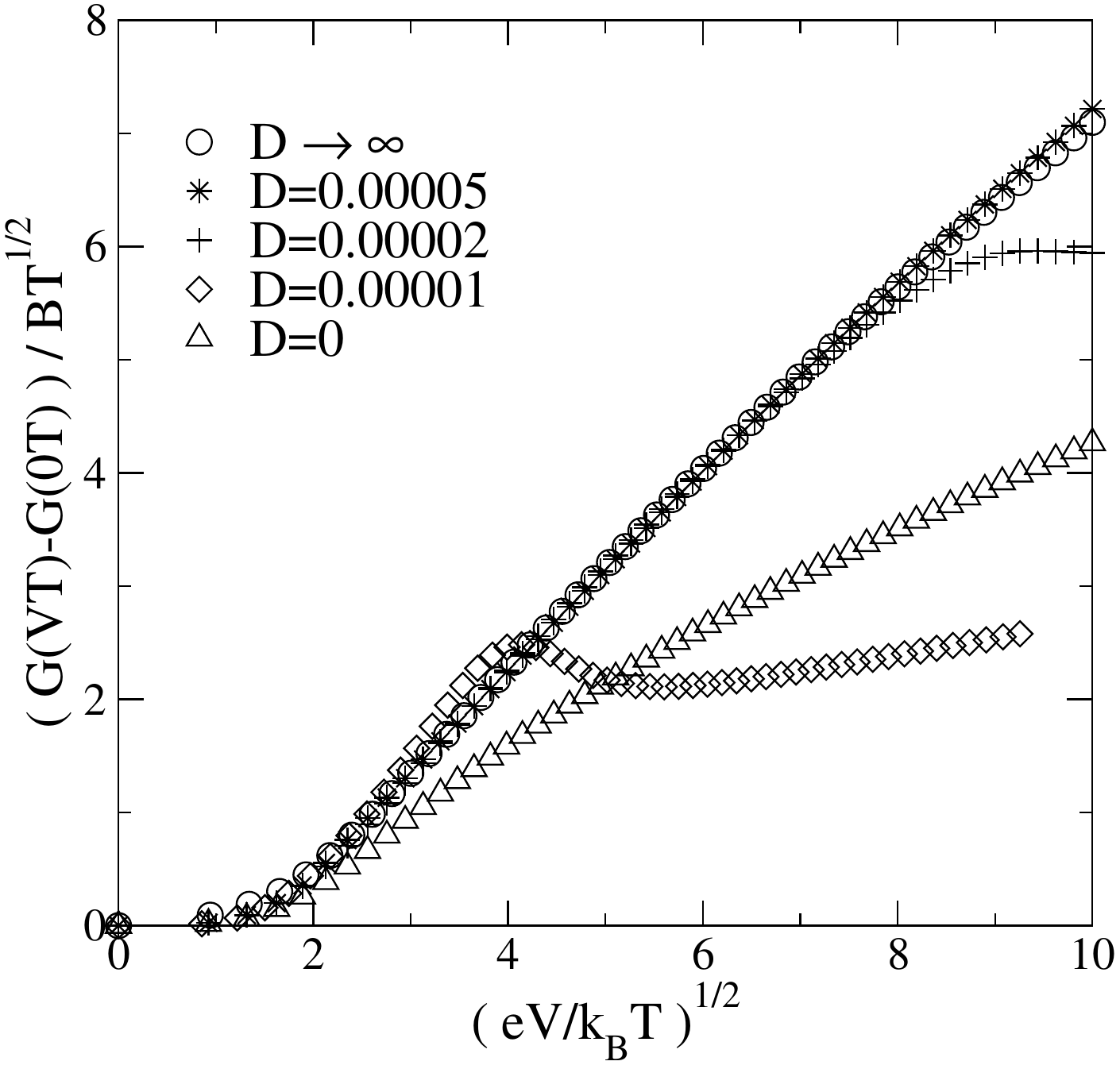}
\caption{Scaling plot of the differential conductance $G(T,V)$ for different values of the anisotropy $D$ at 
a very low temperature, $T = 0.001 T_K$.
Parameters are $W=1$, $E_{32}=-0.67$, $\Delta=0.225$. Here, we define $T_K$ from the half 
width  at the half maximum of the Kondo resonance of the spectral function at equilibrium.}
\label{scaling-vs-d}
\end{figure}

Figure \ref{scaling-vs-d} shows the scaling plot of $G(T,V)$ as a function of $(eV/k_B T)^{1/2}$ 
for a general set of parameters and for several values of the anisotropy $D$. From the half 
width at the half maximum of the Kondo resonance of the spectral function at equilibrium we 
found $T_K^{D=0}\sim 0.00001$ in units of the bandwidth $W$. As it is clear from the figure, when
the anisotropy $D$ becomes $D \ge T_K^{D=0}$ the curves collapse onto a single curve proportional
to $(eV/k_B T)^{1/2}$. 
For $D \ge T_K^{D=0}$ the differential
conductance follows the CFT scaling function expected for the 2CK model. 

\begin{figure}[tbp]
\includegraphics[clip,width=7cm]{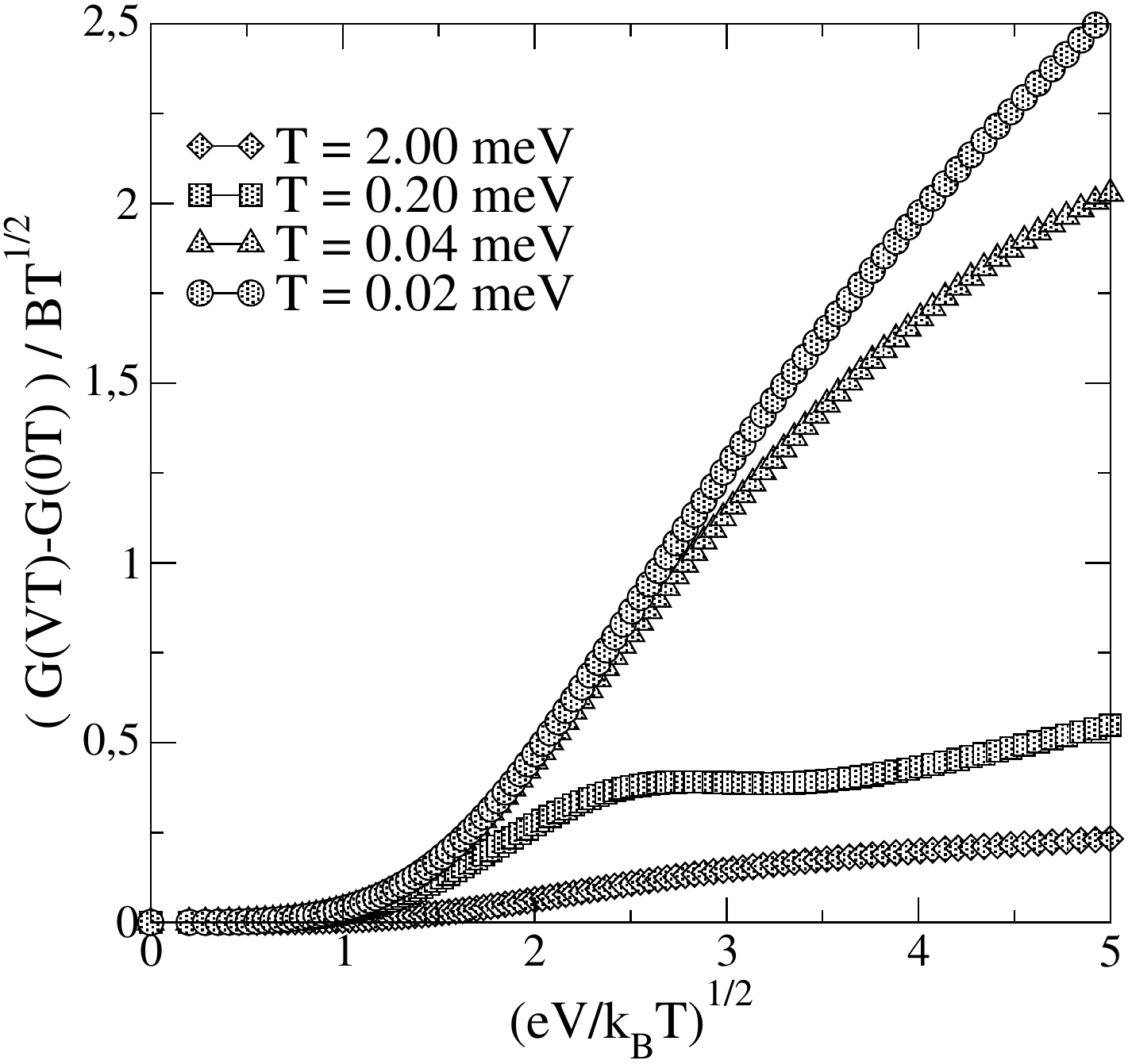}
\caption{Scaling plot of the differential conductance $G(T,V)$ for different temperatures
for the real parameters corresponding to the trimer configuration. }
\label{scaling-vs-t}
\end{figure}

In Fig. \ref{scaling-vs-t} we show the differential conductance as a function of $(eV/k_B T)^{1/2}$ for
the real parameters corresponding to the trimer configuration and for several temperatures. When the 
charge fluctuations are inhibited due to the decreasing temperature, the $G(T,V)$ through 
the Co atom displays the 2C scaling behaviour. 

\section{Conclusions} \label{conclusions}

In summary, extending a previous study of the entropy and equilibrium conductance through a Co
atom coupled to monoatomic Au chains with a four-fold symmetric leads, we have presented 
a comprehensive study of the equilibrium spectral density using NRG and the non-equilibrium conductance
in the non-Ferrmi liquid regime using NCA.

We found that a quantum critical point at the anisotropy value $D=0$ takes place. 
The three different phases, underscreened, overscreened and no Kondo
phases are characterized by the weight of the Kondo resonance at the Fermi level. Within the underscreened
Kondo regime, the value of the spectral density per channel and spin is given approximately by 
$\rho_{\alpha\sigma}(0)\sim\frac{1}{\pi\Delta}$ in analogy to the ordinary Kondo model, although the system is
expected to be a singular Fermi liquid.
On the other hand, within
the overscreened regime, the spectral weight is reduced to half this value, 
$\rho_{\alpha\sigma}(0)\sim\frac{1}{2\pi\Delta}$.
We also found that the 
Kondo temperature of the underscreened phase, $T_K$ plays an important role in the cases of positive
values of $D$. When $0 < D \leq T_K$ the way in which the system enters into the 2C fixed point is 
mediated by a new energy scale $T^\ast$, that depends exponentially on the ratio $T_K / D$. 

We also present a solution of the model by means of the non-crossing approximation not only at 
equilibrium but also for non-equilibrium situations such as transport properties as a function of 
the bias voltage. Our results suggest that the only phase in which the NCA becomes a reliable
method is the overscreened regime in which the anisotropy value should be $D\geq T_K$. In particular,
the NCA for $D=0$ underestimates the spectral weight at the Fermi level in a 40$\%$ and
for the cases in which $0 < D \leq T_K$ the low-energy scale $T^\ast$ is missed.  In contrast,  
for $D \geq T_K$, corresponding to the regime of parameters for which 2C physics is more evident,
we have
verified that the NCA gives the correct residual entropy and (except for a slight overestimation),
it also gives the correct value of the spectral densities at the Fermi level. This suggests that
the NCA is a reliable approximation to study the overscreened regime also at finite bias voltage, for
which our NRG methods are not appropriate.

For small values of 
temperature $T$ and bias voltage $V$, the obtained differential conductance agrees with the predictions of 
conformal-field theory for the 2CK model. Specifically, we show that for realistic parameters 
corresponding to the system of Fig. 1(b)
the conductance through the Co atom as a function of the bias voltage follows a $\sqrt{V}$ dependence, in
agreement with the behavior expected for this kind of non Fermi liquid models. Futhermore a universal scaling 
behaviour as a function of $(eV/k_bT)^{1/2}$ is obtained.

Our results confirm the rich physics of the model. 
We expect that our study can stimulate experimental studies on similar systems.

\section{ACKNOWLEDGMENTS}

This work was partially supported by PIP No 112-200801-01821, 00273 and 00258 of CONICET, 
and PICT 2010-1060 of the ANPCyT, Argentina.
AW acknowledges support from the German Research Foundation
(TR-12, SFB631, NIM, and WE4819/1-1).


\begin{thebibliography}{99}
\bibitem{factor-2} An aditional factor 2 was included due to the definition of the physical operator in 
Eq.(\ref{operador-fisico-aux-part}).
\end{thebibliography}
\end{document}